\documentclass[sigconf]{acmart}
\usepackage[title]{appendix}

\AtBeginDocument{%
  }

\setcopyright{acmlicensed}
\copyrightyear{2024}
\acmYear{2024}
\acmDOI{}

\acmConference[RecSys'24]{ACM Conference on Recommender Systems Workshops}{2024}{Bari, Italy}

\acmBooktitle{ACM Conference on Recommender Systems Workshops, 2024, Bari, Italy}

\begin{document}

\title[Counterfactual Evaluation of Ads Ranking Models through Domain Adaptation]{Counterfactual Evaluation of Ads Ranking Models\\ through Domain Adaptation}

\author{Mohamed A. Radwan}
\affiliation{\institution{Meta}\country{USA}}

\author{Himaghna Bhattacharjee}
\affiliation{\institution{Meta}\country{USA}}

\author{Quinn Lanners}
\affiliation{\institution{Duke University}\country{USA}}

\author{Jiasheng Zhang}
\affiliation{\institution{Meta}\country{USA}}

\author{Serkan Karakulak}
\affiliation{\institution{Meta}\country{USA}}

\author{Houssam Nassif}
\affiliation{\institution{Meta}\country{USA}}

\author{Murat Ali Bayir}
\affiliation{\institution{Meta}\country{USA}}

\renewcommand{\shortauthors}{Radwan et al.}


\begin{abstract}
  We propose a domain-adapted reward model that works alongside an Offline A/B testing system for evaluating ranking models. This approach effectively measures reward for ranking model changes in large-scale Ads recommender systems, where model-free methods like IPS are not feasible. Our experiments demonstrate that the proposed technique outperforms both the vanilla IPS method and approaches using non-generalized reward models.
\end{abstract}




\begin{comment}
\received{20 February 2007}
\received[revised]{12 March 2009}
\received[accepted]{5 June 2009}
\end{comment}

\maketitle

\section{Introduction}
\iffalse
Online advertising has become a crucial component of the global economy, enabling millions of businesses to connect with people anywhere, anytime. With major tech companies investing heavily in this space, platforms such as Instagram, Facebook, YouTube, LinkedIn, and more have become indispensable channels for connecting users with relevant products and services.
\fi

Over the past decade, online advertising has undergone significant transformations, particularly with the integration of Artificial Intelligence and Deep Learning technologies. Leveraging the vast amount of traffic managed by popular recommender services, even minor adjustments to such recommendation models can significantly affect business objectives. Understanding and predicting the impact of such model changes in advance highlights the importance of Offline Evaluation in large scale recommender systems~\cite{Kong2023NeuralInsights}.

While offline evaluation of ranking models can be a trivial task in traditional machine learning setups, it is a challenge in the context of large-scale ad recommendation systems. In such setups multiple models and various system components such as auction mechanisms~\cite{geng_ji_2024}, content selection for organic feed and ads~\cite{biswas2019seeker}, publisher logic, and client-side app logic work together to determine which ads are shown to users. The complexity of these systems significantly affects model-free approaches like Inverse Propensity Scoring (IPS)~\cite{lbottou2012}, as the propensity computation must account not only for the model's output, but also for all other factors that contribute to ad display decisions.

Selection bias is another challenge for offline evaluation of large-scale recommender systems \cite{Chen23, Jadidinejad21}. Offline evaluation uses labeled historical data from the current production model to estimate the performance of a new model that does not have any labeled data from online activity. However, with a large inventory of ads, the new offline model will often rank the ads differently than the current online production model. The distribution difference between the ads presented to users by the current production model and the the ads that the new model would have presented can result in biased offline reward estimates, making it challenging to accurately predict online performance for a new model~\cite{AndurilWorkshop,Weltz2023heteroskedastic}.

Motivated by the challenges outlined above, we propose a new domain-adapted reward model that works on top of an offline A/B testing system to facilitate more accurate ranking model evaluation. Unlike IPS, our approach eliminates the need to account for the complexity of the ads system since the Offline A/B Testing simulation layer handles this aspect. Our experiments demonstrate that the proposed technique outperforms both the vanilla IPS method and basic direct-method reward models.

\section{Methodology}

\subsection{Overview}
\label{section:overview}
\begin{figure}
    \centering
    \includegraphics[width=1\linewidth]{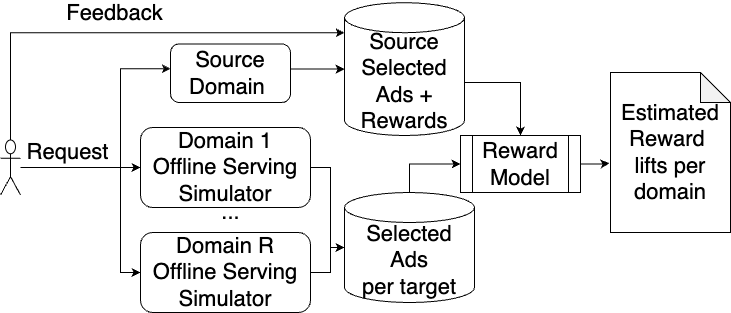}
    \caption{Counterfactual evaluation setup}
    \label{fig:cfe-setup}
\end{figure}

The high level setup for an offline counterfactual evaluation system is shown in Figure~\ref{fig:cfe-setup} . It employs an offline serving simulator that operates a separate instance for each target domain (i.e ads recommended by a given ranking model). The simulator runs recommendation request under each domain. Following simulation, the recommended ads for each corresponding target domain are collected. Subsequently, we leverage the reward model to estimate the expected reward for each domain. In the following sections, we detail how to train such a generalized reward model to effectively handle multiple domains.

\subsection{Notation and set-up}
\begin{comment}
In a typical ad delivery system, the user makes a request to the system which then delivers an ad from a library of candidate ads to based on a policy which is a function of the user, request, and the ads.
\end{comment}

Let $X \subset \mathcal{R}^d$ be a $d$-dimensional context space of user and request features, $A \subset \mathcal{R}^q$ be a $q$-dimensional ad space, and $Y \subset \mathcal{R}$ be a one-dimensional reward space (which may be a discrete set or continuous interval depending on the use case).

Each recommendation model defines a policy which induces a domain in the \( X \times A \) space. We use $S$ and $T_k \in \{T_1, \ldots, T_K\}$ to refer to source and target policies. Finally, we let \( p(y \mid x, a) \) be the conditional distribution of the reward given a particular context and ad combination, which is assumed to be independent of the choice of \( T_k \). The objective is to evaluate a set of K target recommendation models denoted by $T_1$, ..., $T_K$ in an offline setting, using data generated by source policy $S$.

\subsection{Metric}
Our goal is to correctly rank target policies, and therefore we need the reward model to perform fairly across different domains. To achieve this, we choose the coefficient of variance of recovery ($Rec_{cv}$) as the main performance indicator for the reward model:
 \begin{equation}
 {Rec_{cv}} = \frac{Rec_{dev}}{Rec_{avg}}, \quad  
 {Rec(T_k, S)} = \frac{\widehat{\text{Lift}(T_k, S)}}{\text{Lift}(T_k, S)},
\label{eq:Rec}
\end{equation}
where $Rec_{dev}$ is the average absolute deviation of $Rec(.)$, 
and $Rec_{avg}$ is the average of $Rec(.)$ across all target domains. See appendix~\ref{appendix:metrics} for lift definition and additional details.

\subsection{Estimating Lifts Between Domains}
We introduce a model based approach to estimate the lift between two domains. Our approach focuses on modeling the non-overlapping regions between source and target domains and then using the trained reward model to calculate the expected lift of the target domain $T_k$ over the source domain $S$ (for each target domain $k$). Related domain adaptation methods were explored in~\cite{BayirXZS19,Zeng21} within the context of the auction optimization problem. The details of the reward model training are discussed in section \ref{section:training}.

We use $D_{S} = \{(x_i,a_i,y_i)\}_{i=1}^n$, a set of $n$ labeled source domain samples, and $D_{T_k} = \{(x_i,a_i^k)\}_{i=1}^n$, unlabeled target domain data for the same $n$ samples, to estimate the lift as
\begin{equation}
\begin{gathered}
\widehat{Lift(T_k, S)}=
 \\
\frac{1}{n} \left( \sum_{\substack{(x_i, a_i) \in D_{T_{k}, n}}} h(x_i, a_i) - \sum_{\substack{(x_i, a_i) \in D_{S, n} }} h(x_i, a_i) \right),
\label{eq:Lift}
\end{gathered}
\end{equation}
where $h(.)$ is the reward model discussed in the preceding paragraph. Because our offline evaluation setup sends the same requests to all domains, as mentioned in section \ref{section:overview} and shown in Figure~\ref{fig:cfe-setup}, these datasets have the same sample size and sample contexts $x_i$. However, they differ in the ads they recommend, $a_i$, and we only have labels, $y_i$, for the online source recommend model $S$.

\subsection{Reward Model Training}
\label{section:training}
Let \( p_{c}(a \mid x) \) be the probability of observing ad \( a \) under context \( x \) with policy \( c \). We define a weight
\begin{equation}
    w^k_{a} = \frac{p_{T_{k}}(a \mid x)}{p_{S}(a \mid x)}.
    \label{eq:weight}
\end{equation}
$w^k_{a}$ is used in the per sample weights for training a reward model for all $K$ target domains by minimizing the following loss function on the labeled $D_{S}$:

\begin{equation}
\begin{aligned}
\sum_{\substack{(x_i, a_i, y_i) \in D_{S} }} & L[h(x_i, a_i, \theta), y_i] \\
& \times \left[ \sum_{\substack{k = 1}}^K |w_{a_i}^k - 1| + \beta \sum_{\substack{k =1 \\ k' > k }}^K |w_{a_i}^k - w_{a_i}^{k'}| \right],
\label{eq:DsLoss}
\end{aligned}
\end{equation}
where $\theta$ is the model parameters and $\beta$ is a hyper-parameter controlling the contribution of the term  $|w_i^k - w_i^{k'}|$. This term controls the deviation term of Recovery CV by ensuring that the reward model performs equally across all domains, as detailed in Appendix~\ref{appendix:derivation}. 
The $|w_i^k - 1|$ term emphasizes the non-overlapping regions between target and source domains.

\section{Experimental Results}
\begin{figure}
    \centering
    \includegraphics[width=0.5\linewidth]{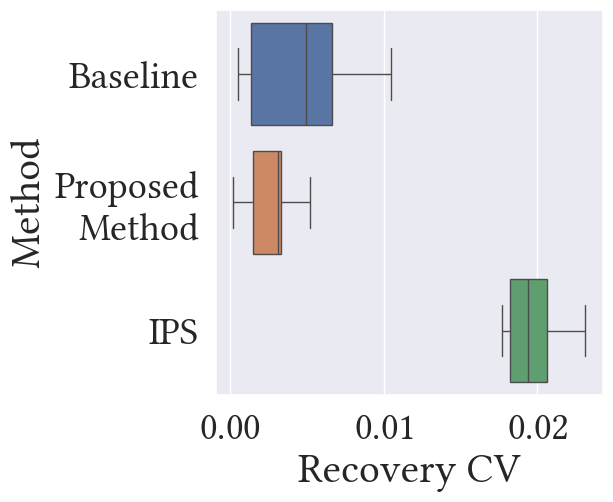}
    \caption{$Rec_{CV}$ of each method used with synthetic data}
    \label{fig:synth_results}
\end{figure}

We report our findings on both synthetic and online experiment results for a CTR prediction model. For the synthetic environment, we generated domain policies, where each target domain or test variant represents an incremental improvement over the source domain or control variant. The reward function is a linear model of the context and ad covariates, where the coefficients are drawn from a normal distribution based on the ad context vector values. We train the baseline solely on the source domain data. We train our proposed reward model using Section~\ref{section:training} weighted target domain information. 
Figure~\ref{fig:synth_results} plots results from three target domains, and shows that our proposed reward outperforms Baseline and IPS.

To evaluate reward models on real experiments, we explored the utilization of a completed A/B test for a CTR prediction model. The test included seven suggested variants that improved upon the control. We used the actual lifts reported per variant as ground truth for evaluating the reward models.  
Given the intractability of propensity score weight in a complex recommendation systems, we train an impression probability estimator per target domain to estimate Eq~\ref{eq:weight} weights. The estimated weights are then used to weight the source data $D_S$, as in Eq~\ref{eq:DsLoss}, to train the proposed reward model. Baseline is trained on $D_S$ without any weighting similar to the synthetic experiment. Our proposed reward model showed a 17.6\% improvement on the Recovery CV metric.



\bibliographystyle{ACM-Reference-Format}
%\bibliography{recsys_bib}


\begin{appendices}
\section{Metric Details}
\label{appendix:metrics}

The metric of interest in our off policy evaluation for ad recommendation models is the performance difference of the offline model with respect to the online source model. We define a lift metric for a target policy $T_k$:
\begin{equation}
Lift(T_k, S) = E_{p_{T_k}}[y] - E_{p_S}[y].
\label{eq:lift_definition}
\end{equation}
The expected reward for target policies cannot be directly estimated from the logged data, which brings us to the core problem: for off-policy evaluation, we want to accurately estimate the lift of a target policy in an offline setting. To reflect this use case, we define a \emph{recovery} metric to compare the performance of the reward model for each target policy $K$,
$Rec(T, T_S) = \frac{\widehat{\text{Lift}(T_k, S)}}{\text{Lift}(T_k, S)}$,
 where \(\widehat{\text{Lift}(T_k, S)}\) is the reward lift estimated by the reward model, and $\text{Lift}(T_k, S)$ is the realised lift . The recovery metric for a domain where the reward model estimates the lift perfectly is $1$. 
 
Moreover, for accurate estimation of policy ranks, we need a metric to perform fairly across all target domains. Therefore, we define an aggregate metric across all target domains using the coefficient of variance:
 \begin{equation}
{Rec_{cv}} = \frac{Rec_{dev}}{Rec_{avg}}, 
\label{eq:rec_cv}
\end{equation}
where 
\begin{equation}
    {Rec_{avg}(T1, T2,.., T_K, S)} = \frac{1}{K} \sum_{T\in [T_1, ..., T_K]} {Rec(T,S)},
\end{equation}
\begin{equation}
    {Rec_{dev}} = \frac{1}{K} \sum_{T\in [T_1, ..., T_K]}{|Rec(T, S) - Rec_{avg}|}.
    \label{eq:rec_dev}
\end{equation}
Here, $Rec_{dev}$ and $Rec_{avg}$ are the standard deviation and average of $Rec$ across all target domains.

\section{Derivation of Recovery Loss}
\label{appendix:derivation}
\subsection{Single-Domain Recovery Optimization}
\label{appendix:single_derivation}
Let us look at the Recovery metric when there is one target domain $T_k$.
Using the definition of recovery from Eq~\ref{eq:Rec} and the definition for lift from Eq~\ref{eq:lift_definition}, we have
$Rec(T_k, S) = 1 - \frac{r_{diff} - \widehat{r_{diff}}}{r_{diff}}$,
where $r_{diff}$ and $\widehat{r_{diff}}$ are the true and estimated difference in rewards between the target and source domains, respectively. Hence the optimization goal is equivalent to minimizing $| r_{diff} - \widehat{r_{diff}} |$.
With domain distributions, the distance can be further expanded as:
\begin{equation}
    \begin{aligned}
    r_{diff} - \widehat{r_{diff}} = (E_{p_{T_k}}[y] - E_{p_S}[y]) - (\widehat{E_{p_{T_k}}[y]} - \widehat{E_{p_S}[y]})
    \\
    = E_{p_{T_k}}[y(a,x) - \widehat{y(a,x)}] - E_{p_{S}}[y(a,x) - \widehat{y(a,x)}]
    \\
    = E_{p_S}\left [ \left (\frac{p_{T_{k}}(a \mid x)}{p_{S}(a \mid x)} - 1 \right ) (y(a,x) - \widehat{y(a, x)}\right ]
    \\
    \approx \frac{1}{N} \times \sum_{(x_i, a_i, y_i) \in D_S} (\frac{p_{T_{k}}(a \mid x)}{p_{S}(a \mid x)} - 1) (y(a,x) - \widehat{y(a, x)}.
\end{aligned}
\end{equation}

It now becomes obvious that the prediction error | $y(a_i, x_i) - \widehat{y(a_i, x_i)}$ | is weighted by $\left (\frac{p_{T_{k}}(a \mid x)}{p_{S}(a \mid x)} - 1\right )$, contributing to the overall distance. In other words, samples $(a_i, x_i)$ with higher $\left (\frac{p_{T_{k}}(a \mid x)}{p_{S}(a \mid x)} - 1\right )$ are more important for optimizing $Rec(.)$.

\subsection{Multi-Domain Optimization}
For the multi-domain use-case, the metric to be optimized is $Rec_{cv}(.)$ defined in Eq~\ref{eq:rec_cv}.
The $Rec_{dev}$ defined in Eq~\ref{eq:rec_dev} can be minimized by minimizing $Rec$ between two target domains $T_k$ and $T_{k'}$:
\begin{equation}
\begin{aligned}
    Rec_{dev} &= \frac{1}{K} \sum_k^K \left| Rec_{T_k} - \sum_{k'}^K \frac{Rec_{T_{k'}}}{K}\right| \\
    &= \frac{1}{K} \sum_k^K \left| \sum_{k'}^K \frac{Rec_{T_k} - Rec_{T_{k'}}}{K}\right| \\
    &\leq \frac{1}{K^2} \sum_k^K \sum_{k'}^K \left| Rec_{T_k} - Rec_{T_{k'}}\right|.
    \end{aligned}
\end{equation}

Furthermore, 
\begin{equation}
\begin{aligned}
| & Rec_{T_k} - Rec_{T_{k'}}| =  \left|\frac{ r_{diff_{t_k}} - \widehat{r_{diff_{t_k}}}}{r_{diff_{t_k}}} - \frac{r_{diff_{t_k'}} - \widehat{r_{diff_{t_{k'}}}}}{r_{diff_{t_{k'}}}}\right| \\
& \approx  \left| \frac{1}{N} \sum_{(x_i, a_i, y_i) \in D_S}^N \left[\frac{1}{r_{diff_{t_{k}}}}\left(\frac{p_{T_{k}}(a \mid x)}{p_{S}(a \mid x)} - 1\right) \right.\right.\\
& \qquad \left.\left.-  \frac{1}{r_{diff_{t_{k'}}}}\left(\frac{p_{T_{k'}}(a \mid x)}{p_{S}(a \mid x)} - 1\right)\right] (y(a,x) - \widehat{y(a, x)} \right|\\
= & \bigg| \frac{1}{N} \sum_{(x_i, a_i, y_i) \in D_S}^N 
 \bigg[\\ &
 \frac{1}{2} \bigg( \frac{p_{T_{k}}(a \mid x)}{p_{S}(a \mid x)} - 1 + \frac{p_{T_{k'}}(a \mid x)}{p_{S}(a \mid x)} - 1\bigg)  \bigg(\frac{1}{r_{diff_{t_{k}}}} - \frac{1}{r_{diff_{t_{k'}}}}\bigg) \\
&   + \frac{1}{2} \bigg(\frac{p_{T_{k}}(a \mid x)}{p_{S}(a \mid x)} -  \frac{p_{T_{k'}}(a \mid x)}{p_{S}(a \mid x)}\bigg)  \bigg(\frac{1}{r_{diff_{t_{k}}}} + \frac{1}{r_{diff_{t_{k'}}}}\bigg)
\\ &  \bigg] \cdot   (y(a,x) - \widehat{y(a, x)} \bigg|
\end{aligned}
\end{equation}

\begin{equation}
\begin{aligned} \label{eq:triangle_ineq_derive}
\le& \bigg| \frac{1}{N} \sum_{(x_i, a_i, y_i) \in D_S}^N \frac{1}{2}  \bigg[ 
\\ & \bigg|\bigg(\frac{p_{T_{k}}(a \mid x)}{p_{S}(a \mid x)} + \frac{p_{T_{k'}}(a \mid x)}{p_{S}(a \mid x)} - 2\bigg)\bigg(\frac{1}{r_{diff_{t_{k}}}} - \frac{1}{r_{diff_{t_{k'}}}}\bigg)(y(a,x) - \widehat{y(a, x)}\bigg| 
\\ & + \bigg|\bigg(\frac{p_{T_{k}}(a \mid x)}{p_{S}(a \mid x)} - \frac{p_{T_{k'}}(a \mid x)}{p_{S}(a \mid x)}\bigg)\bigg(\frac{1}{r_{diff_{t_{k}}}} + \frac{1}{r_{diff_{t_{k'}}}}\bigg)(y(a,x) - \widehat{y(a, x)}\bigg| 
\bigg]\bigg|
\end{aligned}
\end{equation}

Eq~\ref{eq:triangle_ineq_derive} follows from the triangle inequality. From it we can deduce that $\left| Rec_{T_k} - Rec_{T_{k'}}\right|$ can be minimized by minimizing two separate terms. The first is:

\begin{equation}
\begin{aligned}
\frac{1}{N} \sum_{(x_i, a_i, y_i) \in D_S}^N \frac{1}{2} &\left(\frac{p_{T_{k}}(a \mid x)}{p_{S}(a \mid x)} -1 + \frac{p_{T_{k'}}(a \mid x)}{p_{S}(a \mid x)} - 1\right)
\\ &\left(\frac{1}{r_{diff_{t_{k}}}} - \frac{1}{r_{diff_{t_{k'}}}}\right)(y(a,x) - \widehat{y(a, x)},
\end{aligned}
\end{equation}

which is similar to reducing the recovery term for each domain as discussed in Appendix~\ref{appendix:single_derivation}. The second term,
\begin{equation}
\begin{aligned}
\frac{1}{N} \sum_{(x_i, a_i, y_i) \in D_S}^N \frac{1}{2} &\left(\frac{p_{T_{k}}(a \mid x)}{p_{S}(a \mid x)} - \frac{p_{T_{k'}}(a \mid x)}{p_{S}(a \mid x)}\right)
\\ &\left(\frac{1}{r_{diff_{t_{k}}}} + \frac{1}{r_{diff_{t_{k'}}}}\right)(y(a,x) - \widehat{y(a, x)},
\end{aligned}
\end{equation}

emphasizes the influence of the error from samples whose ad impression probability is different between target domains. We use these two weights to derive the sample weights for our Eq~\ref{eq:DsLoss} loss function:
\begin{equation}
\begin{aligned}
\sum_{\substack{(x_i, a_i, y_i) \in D_{S} }} & L[h(x_i, a_i, \theta), y_i] 
\times \bigg[ \sum_{\substack{k = 1}}^K \left|\frac{p_{T_{k}}(a \mid x)}{p_{S}(a \mid x)} -1\right| + \\& \beta \sum_{\substack{k =1 \\ k' > k }}^K \left|\frac{p_{T_{k}}(a \mid x)}{p_{S}(a \mid x)} - \frac{p_{T_{k'}}(a \mid x)}{p_{S}(a \mid x)}\right| \bigg].
\end{aligned}
\end{equation}

\end{appendices}

\end{document}